\documentclass{article}
\usepackage{graphicx}

\date{~}

\begin{document}

\title{%
\vskip-6pt \hfill {\rm\normalsize UCLA/01/TEP/3} \\
\vskip-6pt \hfill {\rm\normalsize CWRU-P4-01} \\
\vskip-6pt \hfill {\rm\normalsize  June 2001} \\
\vskip-9pt~\\
Detectability of a subdominant  density component of cold dark matter}

\author{Gintaras Duda\rlap{,}{$^{1}$} Graciela Gelmini\rlap{,}{$^{1}$}
Paolo Gondolo\rlap{$^{2}$}
  \\ ~\\
  \small \it ${}^{1}$ Dept. of Physics and Astronomy, UCLA (University of
  California, Los Angeles),\\
  \small \it 405 Hilgard Ave., Los Angeles, CA 90095, USA\\
  \small \it {\rm gkduda,gelmini@physics.ucla.edu}
  \\~\\
  \small \it ${}^{2}$ Dept. of Physics, Case Western Reserve University,\\
  \small \it 10900 Euclid Ave, Cleveland, OH 44106-7079, USA\\
  \small \it {\rm pxg26@po.cwru.edu}
  \\~\\
  }

\maketitle

\begin{abstract}
  Here we examine the detectability of collisionless dark matter candidates
  that may constitute not all but only a subdominant component of galactic cold
  dark matter. We show that current axion searches are not suited for a
  subdominant component, while direct WIMP searches would not be severely
  affected by the reduced density. In fact, the direct detection rates of
  neutralinos stay almost constant even if neutralinos constitute 1\% of the
  halo dark matter. Only for lower densities do the rates decrease with
  density.  Even neutralinos accounting for only $10^{-4}$ of the local dark
  halo density are within proposed future discovery limits. We comment also on
  indirect WIMP searches.
\end{abstract}

\maketitle

%\section{Introduction}

Claims for the need of collisional cold dark matter \cite{NCCDM} as the
main form of dark matter in the Universe have led us to consider the
observability of collisionless cold dark matter (CCDM) when it is merely 
a subdominant component of the cold dark matter (CDM). 
Namely, if the previously favored CDM candidates, such as axions
or Weakly Interacting Massive Particles (WIMPs), constitute only a fraction, say
1\% or less, of the local dark matter
density, would these particles still be observable in the current and proposed
direct and indirect dark matter searches?  This is a valid question even if
non-CCDM is proven not to be necessary. In fact there is always the possibility
of the CDM consisting of several populations, the one we are searching for not
being the dominant one. We could even reverse our question in the following
manner. If we see a CDM signal in any of our searches, could we be observing a
subdominant component of the total CDM?

Naively one may claim that if the local CDM density is 1\%, say, of the local
halo density, the expected rates in CDM detectors, being proportional to the
local number density, should decrease by the same amount. However, we note that
a reduction in the relic CDM density implies in general an increase in the
probability of interaction of CDM with the detector, for example an increase in
the WIMP--nucleus cross section or an increase in the axion--photon coupling
constant. Since the detection rate depends on the product of the interaction
probability and the local CDM density, the increase in interaction probability
 may compensate
the decrease in CDM density, and the detection rate would remain unchanged.

For axions, this argument is new; for WIMPs, it is not. It has been mentioned
implicitly or explicitly in many papers on WIMP detectability since the
inception of the subject\cite{MANY}. It is timely, we believe, to pinpoint,
emphasize and update this argument, because it clearly points to the value of
continuing WIMP searches even if WIMPs constitute only a small fraction of the
dark matter.

We now present arguments that the compensation between interaction probability
and local density occurs for axions and WIMPs, and point out some exceptions.

Unless there is segregation for different types of dark matter, the ratio of
CCDM to total DM should be the same locally in the Galaxy and globally in the
whole Universe.  Thus in the following we assume that the local fraction of
CCDM $f_{CCDM}$ is related to the CCDM relic density $\Omega_{CCDM}$ through
\begin{equation}
\label{no-segregation}
f_{CCDM} = \frac{\rho_{CCDM}}{\rho_{\rm local}} =
\frac{\Omega_{CCDM}}{\Omega_{DM}},
\end{equation}
where $\rho_{CCDM}$ is the local density of a particular CCDM candidate,
$\rho_{\rm local}\simeq$ 0.3 GeV/cm$^3$ is the local halo density (at the
location of the Earth), $\Omega_{CCDM}$ is the relic density of our particular
CCDM candidate, and $\Omega_{DM} \simeq$ 0.3 is the total contribution of DM to
the total energy density of the Universe.

Because the relic density of axions is directly related to its mass, and axion
searches are tuned to the axion mass, current searches are not suited to
look for a subdominant axion component. The axion relic density is directly
related to its mass $m_a$.  The usual
relation (which has its caveats, see for example \cite{Sikivie} and references
therein) between the axion relic density and its mass $m_a$ is, for a
QCD constant of 200 MeV,
\begin{equation}
\label{omega_a}
m_a \simeq \frac{0.6 \times 10^{-5} {\rm eV}}{(\Omega_a h^2)^{\frac {6}{7}}},
\end{equation}
where $h$ is the reduced Hubble constant, $h\simeq 0.7$. A dominant component
of axions with $\Omega_a= 0.3$ corresponds, according to this relation, to $m_a
=3 \times 10^{-5}$eV. Thus, we could decrease the density at most to $\Omega_a=
0.003$, so that axions contribute 1\% of the total DM density, before
encountering the upper bound of $3\times 10^{-3}$eV on the axion mass derived
from the observed duration of the Supernova 1987A neutrino signal (and other
bounds which exclude all heavier axions, see for example \cite{Sikivie} and
references therein).

The power $P$ from axion to photon conversion in an electromagnetic cavity
used for axion dark matter searches is proportional to the product $\rho_a m_a$
of the local axion density and the axion mass \cite{Sikivie}. In absence of
segregation, eq.~(\ref{no-segregation}) shows that the power is also
proportional to $\Omega_a m_a$, which using eq.~(\ref{omega_a}) for the axion
relic density gives
\begin{equation}
  P \propto \Omega_a^{1/7} ,
\end{equation}
that is the power is proportional only to the 1/7th power of the axion relic
density. For a decrease in $\Omega_a$ by a factor of 100, the power decreases
only by a factor of 2. Of course, because the axion mass has shifted to keep
relation (\ref{omega_a}) valid, this power is now at a frequency which is 500
times larger and one would need resonant cavities consequently smaller.
The limiting factor of  axion dark matter searches with electromagnetic cavities
is not the axion to photon conversion power, but the size of the necessary
cavities.

The relic density of WIMPs $\Omega_\chi$ is determined by their
annihilation cross section $\sigma_a$ by the relation
\begin{equation}
\Omega_{\chi} h^2 \simeq \frac{1 \times 10^{-37} {\rm cm}^2}
{\langle \sigma_a v \rangle}~,
\end{equation}
where $\langle \sigma_a v\rangle$ is the thermal average of the annihilation
cross section times the relative velocity of the WIMPs at freeze-out.  A
reduction in the relic WIMP density requires an increase in their annihilation
cross section in the early Universe. This increase is often
associated with an increase in the scattering cross section $\sigma_s$ of WIMPs
off atomic nuclei. Since the interaction rate in detectors depends on the
product $\sigma_s \rho_{\chi}$, if the scattering cross section increases as
much as the annihilation cross section, the rate would be unchanged even if
$\rho_{\chi}$ has decreased. Concerning indirect detection, the flux of rare
cosmic rays and of gamma-rays produced in halo annihilations depends on the
product of the square of the density and the annihilation cross section into a
particular channel, $\sigma_a {\rho_\chi}^2$. Thus, even if an increase
in the cross section would compensate the decrease in one of the powers of the
density, the fluxes would still decrease linearly with the  halo WIMP density.
However, the intensity of the high-energy neutrino emission from the Sun and the 
Earth
would in many cases decrease only slightly, because, to the extent
that capture and
annihilation of WIMPs in the Sun and the Earth have the time to equilibrate,
the neutrino intensity depends only on the capture rate which in turn depends
on the product $\sigma_s \rho_\chi$.

We can understand the relation between the scattering and annihilation cross
sections $\sigma_s$ and $\sigma_a$ as follows. The scattering cross section of
a WIMP of mass $m_{\chi}$ with a nucleus of mass $m_N$ is of the form
\begin{equation}
\sigma_s  \simeq \frac{m_{\chi}^2 m_N^2}{(m_{\chi} + m_N)^2} |A_s|^2~,
\end{equation}
where $A_s$ is a reduced amplitude which depends on the dynamics of the
collision. The annihilation cross section of WIMPs into light particles is
\begin{equation}
\sigma_a  \simeq N_a m_{\chi}^2 |A_a|^2~,
\end{equation}
where $A_a$ is the corresponding reduced amplitude and $N_a$ is the number of
annihilation channels. In the case of interactions of weak order, the
amplitudes are of the order,
\begin{equation}
|A_a|^2  \simeq \frac{\alpha^2}{M^2}~,~~~~~~ |A_s|^2  \simeq A^2
\frac{\alpha^2}{M^2}~,
\end{equation}
where $\alpha$ is a coupling constant of weak order $\alpha \simeq 10^{-2}$,
$M$ is a mass of the particles mediating the interaction, typically $M \simeq
100$GeV and $A$ is the atomic number of the interacting nucleus. Our expression
for the scattering amplitude includes the nuclear coherent enhancement factor
$A^2$  valid for spin-independent scattering; for
spin-dependent scattering the factor $A^2$ should be dropped.  Also, our
expression for the annihilation cross section is valid for $m_{\chi} < M$,
while in the opposite range, $m_{\chi} > M$, we expect $\sigma_a \simeq N_a
m_{\chi}^{-2}$.

The simplest case to consider is that of WIMPs lighter than the nuclei they
interact with. From the above equations it is obvious that for these WIMPs
\begin{equation}
\frac{\sigma_s} {\sigma_a}  \simeq \frac{|A_s|^2}{|A_a|^2} \simeq {\rm const}
\end{equation}
the ratio of cross sections is approximately constant. In fact, provided the
main annihilation channel is into fermions, quarks in particular, crossing
arguments insure that the reduced amplitudes of annihilation and scattering
with nucleons are similar.

Heavier WIMPs may have other annihilation channels, such as Higgs bosons or
vector boson pairs. The crossing argument then does not apply and we don't
expect the scattering amplitude to grow as much as the annihilation amplitude.
Moreover, for WIMPs heavier than the nuclei they scatter from, the scattering
cross section becomes largely independent of the WIMP mass, while the
annihilation cross section always depends on $m_\chi$. In this case, while the
annihilation cross section could be made larger by considering lighter (if
$m_{\chi} > M$) or heavier (if $m_{\chi} < M$) WIMPs, the scattering cross
section would remain largely unchanged.

Therefore, for relatively light WIMPs, and to a lesser extent for heavy WIMPs,
we expect the scattering cross section to grow by the same factor
$\Omega_{DM}/\Omega_{\chi}$ the annihilation cross section needs to grow to
reduce the local CDM density by $\Omega_{\chi}/\Omega_{DM}$. So the rate, which
is proportional to the product of the local CDM density and the scattering
cross section, remains unchanged.

This argument ceases to be applicable at some small enough WIMP densities,
because the necessary increase in cross sections is due to larger couplings
and/or smaller mediator masses, which, at some point, encounter accelerator
limits which exclude the model. In fact Fig.~1 (described below)  shows that for
neutralinos constituting   10\% of the halo or more the direct detection rates
are largely maintained (as evidenced by the behavior of the envelope of the
highest rates), and for densities as low as 1\% of the halo density,
the highest rates only decrease by a factor of about three,
showing that there is compensation in the
interaction rates while densities decrease by a factor of up to 100. As
mentioned, the compensation ceases to work for smaller densities, and for these
(as can be seen in Figs.~1 and 2) the envelope of highest rates decreases
linearly with the density.

To substantiate the general arguments presented so far, we have analyzed the
concrete case
of the lightest neutralino in usual variations of the Minimal Supersymmetric
Standard model. We used a table of models allowed by all accelerator limits,
produced with the DarkSUSY code \cite{DarkSUSY} over the last few years for
other purposes, i.e. having in mind other issues which were addressed in the
papers of Ref \cite{DarkSusypapers} for which the models were originally
computed. We have, therefore, not done any particular sampling of the models to
favor lower densities and higher detection rates.  We restricted our attention
to models with $\Omega_\chi \leq \Omega_{DM}= 0.3$ ($\Omega_\chi
 h^2 \leq 0.15$) for which we found about
45,000 points in parameter space. For these models, using the spin-dependent and
spin-independent neutralino-nucleon cross sections provided in the table, we
computed the integrated interaction rates on Ge, following L. Bergstr\"om
and P. Gondolo in ref. \cite{DarkSusypapers}. We plot the resulting integrated
rates (in units of events per kg-day) in the first two figures of this paper.

Figs. 1 and 2 show the expected integrated rates  in Ge detectors as
function of the lightest neutralino relic density. Fig.~1 shows only a part of
Fig.~2 (the part with the highest rates and densities) displaying the original
points in the table of models.  Fig.~2 shows the whole range of densities (which
reach up to $\Omega_\chi h^2 \simeq 10^{-6}$) using a regular grid of
points covering the region with models.

In Fig.~1 the change of the slope of the envelope of the points with
maximal rate as the density diminishes is clearly evident.  There is
approximately no change in maximal rates in the first decade of decrease of
density, from $\Omega_\chi h^2=0.15$ (for which neutralinos constitute the
whole halo, $f_{CCDM}=1$) to 0.015 (for which neutralinos constitute 10\% of
the halo, $f_{CCDM} = 0.1$).  There is only about a factor of 3 decrease in the
next decade, from $\Omega_\chi h^2=0.015$ to 0.0015 ($f_{CCDM}$ from 0.1 to
0.01). For smaller densities the slope of the envelope clearly changes, and as
evidenced by Fig.~3, the maximal rates decrease linearly with $\Omega_\chi h^2$
up to the smallest densities. Some of the points shown in Figs.~1 and 2,
mostly among with the smallest densities in Fig.~3, should correspond to
resonances in the annihilation cross section.

The compensation in the rates can be largely understood just by looking at the
spin-independent neutralino-proton cross section $\sigma_{\chi-p}$ as a
function of the lightest neutralino relic density, shown in Fig.~3, again with
a regular grid showing the allowed region where points were found. Also from
this figure, looking at the envelope of the highest cross sections, it is
evident that for $\Omega_\chi h^2$ decreasing from 0.15 to 0.0015, i.e. in the
first two decades of decrease in neutralino density, $\sigma_{\chi-p}$
increases with decreasing densities; this leads to a compensation in the direct
rates. On the other hand, for smaller densities, $\sigma_{\chi-p}$ is about
constant or decreases slightly with decreasing densities; this effect is due to
accelerator bounds. 

Since experimental upper bounds and discovery regions are at present given in
terms of $\sigma_{\chi-p}$, Fig.~3 shows the approximate level of the claimed
signal and present bounds (by the DAMA, CDMS, COSME-IGEX, and
Heidelberg-Moscow  collaborations \cite{DAMAetc}) and conceivable future
discovery level (by the GENIUS proposal \cite{GENIUS}) which are  of order 
$10^{-5}$
pbarns and $10^{-9}$ pbarns, respectively, for neutralinos which
account for the whole local halo density, i.e. with $f_{CCDM}$=1. (These values
depend on the neutralino mass, but to simplify the presentation we only take the
most conservative bounds in our range of masses.) In our case
these values must be understood as levels of $f_{CCDM}  \sigma_{\chi-p}$, which
are shown in Fig.~3 (with short-dashed and long dashed lines respectively).  The
present level of discovery
lightly touches the boundary of the highest rates for densities reduced by up to
a factor of about 10. This suggests the possibility that the DAMA claimed signal
may correspond to subdominant neutralinos. It is very interesting to see
that many models of subdominant neutralinos even  with $10^{-4}$ of the
total dark matter density, enter in the discovery limit proposed by Genius.

In conclusion, the main point of this paper is that the direct detection 
rates of neutralinos remain about constant for  neutralino densities 
 between 100\% and  1\% of the halo dark matter and only decrease linearly with 
the 
density for lower densities. Thus if a signal is found in direct detection 
experiments the question of which component of dark matter was found,
the primary or a sub-dominant one, may remain open. We also note 
that neutralinos with density as small as $10^{-4}$ of the
local dark halo density are within the discovery limits of proposed future
experiments.

\section*{Acknowledgments}

G.D. and G.G. were supported in part by the U.S. Department of Energy Grant No.
DE-FG03-91ER40662, Task C.  P.G.\ acknowledges that one of the points made in
this paper, namely that in the MSSM some extreme low density neutralinos with
$\Omega h^2 \simeq 10^{-3}-10^{-5}$ might be detectable, was made in
conversations with Joe Silk and Joakim Edsj\"o in March 1999.

\newpage

\section*{Figure Captions}

\begin{description}

\item[Fig. 1] Integrated interaction rates of neutralinos in Ge detectors 
(computed as in  L. Bergstr\"om  and P. Gondolo Ref. \cite{DarkSusypapers}) in 
units of events per kg-day, as function of the neutralino relic density, for 
$\Omega_\chi h^2 \leq 0.15$. Each point represents an actual model.
 
\item[Fig. 2] Integrated interaction rates of neutralinos on Ge 
extended to the whole 
range of densities. A regular grid of points shows the region covered with 
models.

\item[Fig. 3] Spin-independent neutralino-proton cross section $\sigma_{\chi-p}$ 
as function of the lightest neutralino relic density. As in Fig.2, a regular 
grid of points shows the region where models were found. The short-dashed and 
long dashed lines of $f_{CCDM}  \sigma_{\chi-p}$=$10^{-5}$ pbarns and $10^{-9}$ 
pbarns show the approximate level of DAMA claimed signal and the current bounds, 
 and the  conceivable future discovery level, respectively ($f_{CCDM}$
is the fraction of the local halo density consisting of neutralinos).

\end{description}

\newpage
\begin{figure}[t]
\includegraphics[width=\textwidth]{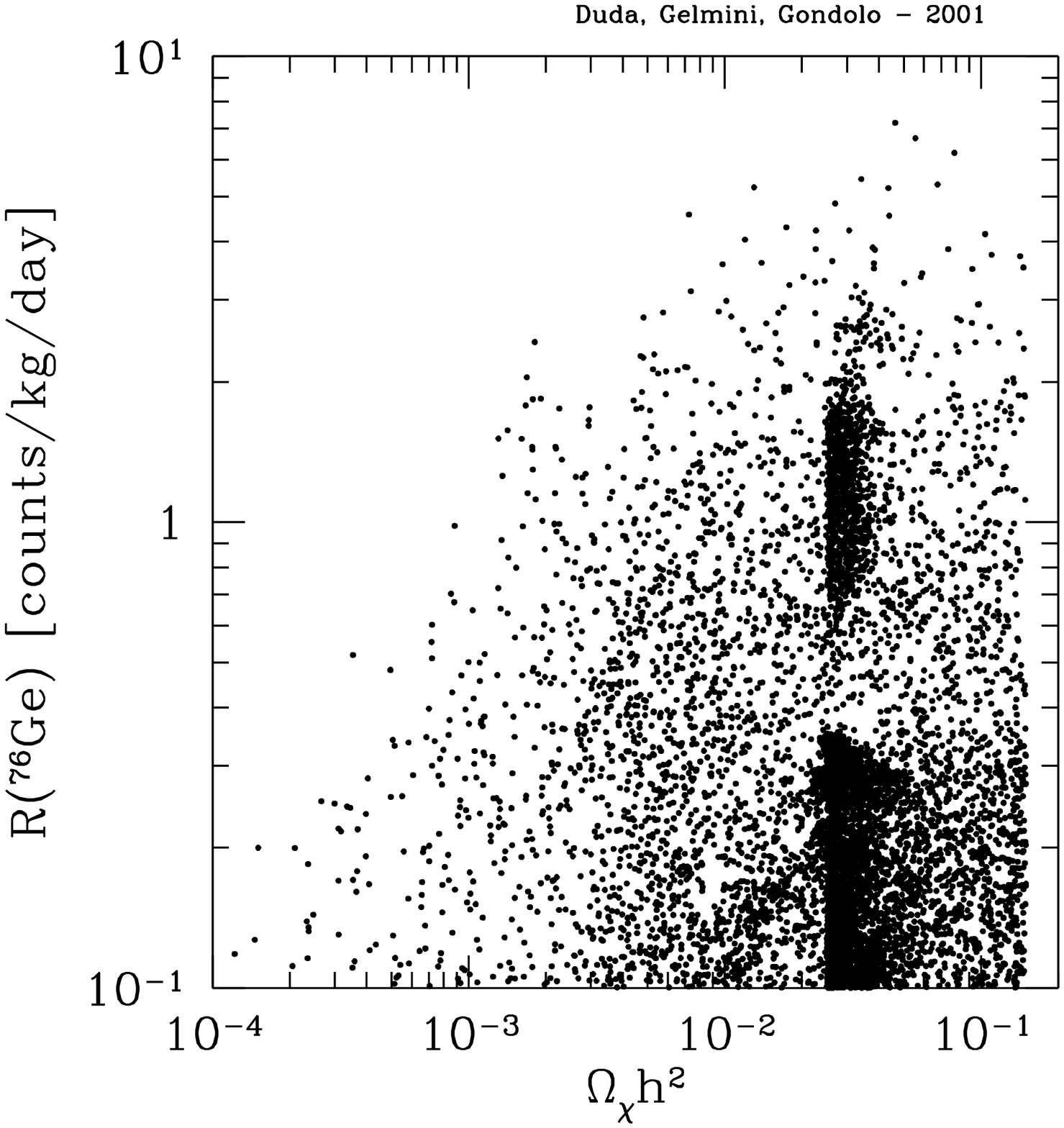}
Figure 1.
\end{figure}

\newpage
\begin{figure}[t]
\includegraphics[width=\textwidth]{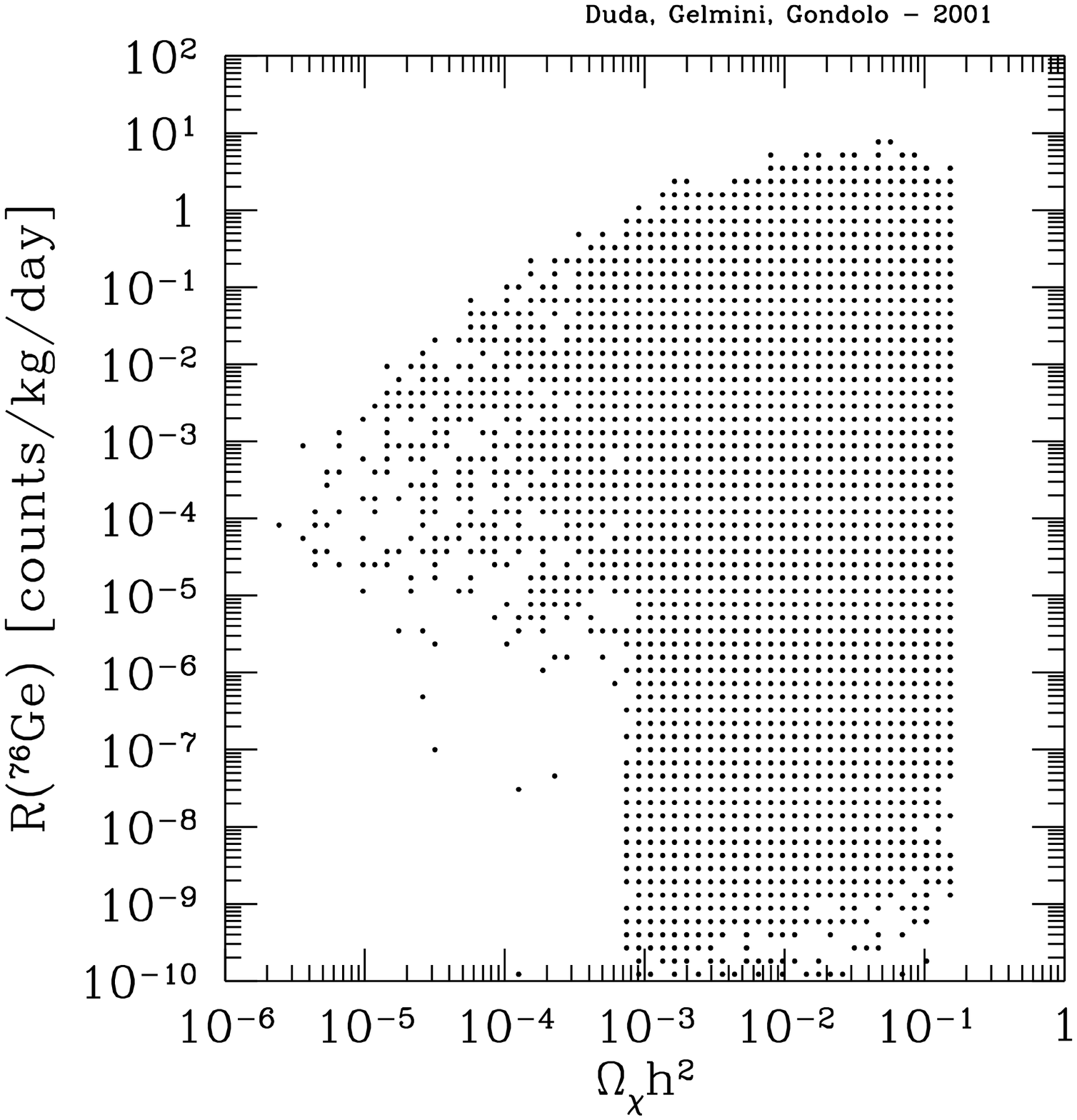}
Figure 2.
\end{figure}

\newpage
\begin{figure}[t]
\includegraphics[width=\textwidth]{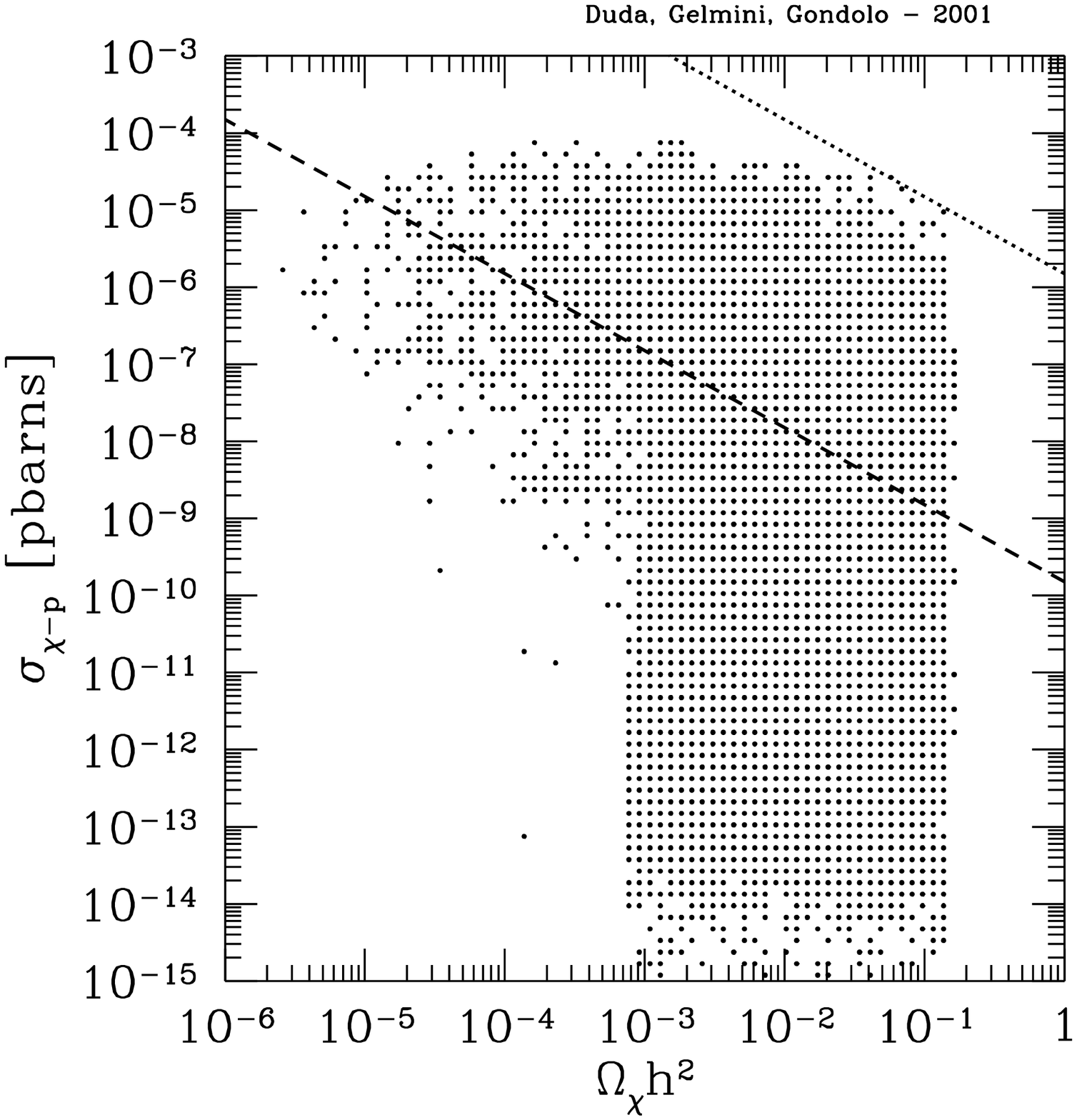}
Figure 3.
\end{figure}


\begin{thebibliography}{99}

\bibitem{NCCDM}

D.~N.~Spergel and P.~J.~Steinhardt,
Phys.\ Rev.\ Lett.\ {\bf 84}, 3760 (2000)

\bibitem{MANY}
  
  See, for example, T.~K.~Gaisser, G.~Steigman and S.~Tilav, Phys.\ Rev.\ D
  {\bf 34}, 2206 (1986); B. Sadoulet, in Proc. of the ``13th Texas Symposium on
  Relativistic Astrophysics", Chicago, Illinois, Dec. 1986, p. 260; K. Griest
  and B. Sadoulet, in Proc. of the ``Second Particle Astrophysics School on
  Dark Matter", Erice, Italy, 1988; G.~Gelmini, E.~Roulet and P.~Gondolo,
  Nucl.\ Phys.\ Proc.\ Suppl.\ {\bf 14B}, 251 (1990) and \ Phys.\ B {\bf 351},
  623 (1991); A.~Bottino et al., Astropart.\ Phys.\ {\bf 2}, 77 (1994);
  F.~Halzen, in ``Int. Symp. on Particle Theory and Phenomenology", Ames, Iowa,
  May 1995, astro-ph/9508020; P.~Gondolo, in XXXI Rencontre de Moriond ``Dark
  Matter in Cosmology, Quantum Measurements, Experimental Gravitation'', Les
  Arcs, France, January 1996, astro-ph/9605290; V.~Berezinsky et al.
  Astropart.\ Phys.\ {\bf 5}, 1 (1996); L. Bergstr\"om and P. Gondolo,
  Astropart.\ Phys.\ {\bf 5}, 183 (1996); A.~Bottino, F.~Donato, N.~Fornengo
  and S.~Scopel, Astropart.\ Phys.\ {\bf 13}, 215 (2000) and Phys.\ Rev.\ D
  {\bf 63}, 125003 (2001).


\bibitem{Sikivie}

 P. Sikivie, hep-ph/0002154 and references therein.

\bibitem{DarkSUSY}

  P. Gondolo, J. Edsj\"o, L. Bergstr\"om, P. Ullio, and E.A. Baltz,
  in preparation; {\it ditto},
  astro-ph/0012234, to appear in the proceedings of the 3rd International
  Workshop on the Identification of Dark Matter (IDM2000) in York.

\bibitem{DarkSusypapers}

  L. Bergstr\"om and P. Gondolo, Astropart.\ Phys.\ {\bf 5}, 183 (1996); J.
  Edsj\"o and P. Gondolo, Phys.\ Rev.\ {\bf D56}, 1879 (1997); L.  Bergstr\"om,
  P. Ullio, and J. H. Buckley, Astropart.\ Phys.\ {\bf 9}, 137 (1998); L.
  Bergstr\"om, J. Edsj\"o, P. Gondolo, Phys.\ Rev.\ {\bf D58}, 103519 (1998);
  E.A.  Baltz and J. Edsj\"o, Phys.\ Rev.\ {\bf D59}, 023511 (1999).


\bibitem{DAMAetc} R. Bernabei et al. [DAMA col.],   Phys. Lett. {\bf B 480}, 23
(2000); R. Abusaidi et al. [CDMS col.], Phys. Rev.  Lett. {\bf 84}, 5699 (2000);
I.G. Irastorza et al. [COSME-IGEX col.] hep-ph/0011318; L. Baudis et al.
[Heidelberg-Moscow col.]  Phys. Rev. {\bf D59}, 022001 (1999).

\bibitem{GENIUS}
 L. Baudis et al. Phys. Rep. {\bf 307}, 301 (1998).

\end{thebibliography}
\end{document}